# IS SCHROEDINGER EQUATION CONSISTENT WITH INFORMATION THEORY?

R. P. Venkataraman* , 1371, 13th Main Second Stage, First Phase, B.T.M. Layout, Bangalore – 560 076

ABSTRACT  It is shown that Schroedinger equation is not consistent with information theory. From the modified form of information which ensures that the most probable density function it yields tallies with a general form of continuous Riemann integrable density function that has real or imaginary zeros or singularities at end points of *[a,b]∈R*, a new variational formulation for quantum mechanics is proposed that yields a system of Euler-Lagrange equations that are non-linear. It is proved that the solutions of this system are unique, orthonormal and complete. One dimensional harmonic oscillator has been solved.

## I INTRODUCTION

Quantum mechanics makes probabilistic predictions and information theory (Shannon et al, 1969 and Jones, 1979) yields the most probable density function subject to a given set of constraints (Venkataraman, 1999). If the probabilistic predictions of quantum mechanics (Messiah, 1968 and Schiff, 1968) are also the most probable ones, then Schroedinger equation will be consistent with information theory. An affirmative answer to the title question would then render a comparative study of an ensemble of classical systems, generated, for instance, by replacing the initial conditions of classical equations with random variables, with quantum system meaningful. In section II the question is answered in the negative and a new variational formulation of quantum mechanics is proposed in section III. The latter involves modified form of information (Venkataraman, 1999) that ensures the probability density function of a real variable obtained from it tallies with a general form of continuous Riemann integrable density function in *[a,b]* that has real or imaginary zeros or singularities at end points. In appendix 1 multivariate probability densities and, convergence of power series in several variables using spherical polar coordinates are discussed. In appendix 2 it is proved that for the proposed variational formulation the Lagrange multipliers corresponding to orthogonality are zero and that the orthonormal set of solutions is unique and complete in the corresponding Hilbert space. In section IV the system of non-linear Euler-Lagrange equations is solved for one dimensional harmonic oscillator. Results are summarised in section V.

## II. Information Theory, Probability and Schroedinger Equation

Information theory predicts the most probable density function subject to a given set of constraints. If *ρ(x)dx* is the probability density in *[a,b]⊂R*, the information *I* associated with it is given

$$I = <\ln \rho> \qquad (1)$$

where the brackets denote the averaging done with respect to *ρ(x)dx*. If the constraints are

$$<x^i> = c_i \qquad (2)$$

---
* rpvraman@yahoo.com

where $c_i$'s are constants, then $\rho(x)dx$ is obtained by minimizing $I$ subject to the given linearly independent set of constraints. The variational method yields

$$\rho(x)dx = \exp(-\Sigma_{i=0}^{\infty} a_i x^i)dx \qquad (3)$$

where $a_i$'s are Lagrange multipliers to be determined from

$$\partial/\partial a_i \langle -\ln \rho(x) \rangle = c_i \qquad (4)$$

Schroedinger equation can also be derived from the variational principle of minimizing $\langle H \rangle$ where $H$ is the Hamiltonian subject to normalisation of eigenvectors. $\langle \ln\rho(x) \rangle$ does not occur in the variational formulation and the inference thereof is that Schroedinger equation is not consistent with information theory or that the predictions of quantum mechanics are not the most probable ones.

Before providing an alternative variational formulation of quantum mechanics, a natural question to ask is "Is the probability density function specified in equation (3) by fixing all moments the most general one in *[a,b]* ? " If $\rho(x)dx$ is continuous and has no real or imaginary zeros or singularities in *[a,b]* Taylor expansion of $\ln\rho(x)$ about $x_0 \in [a,b]$ yields

$$\ln\rho(x) = \ln\rho(x_0) + \Sigma_{i=1}^{\infty}(x-x_0)^i 1/i! [\frac{d^i}{dx^i}\ln\rho(x)]_{(x=x_0)} \qquad (5)$$

From the above equation, equation (3) follows, as its R.H.S. is convergent. Hence equation (3) represents a continuous probability density function that has no zeros or singularities in *[a,b]*. If $\rho(x)dx$ has real or imaginary zeros or singularities at end points of *[a,b]* specified by $Z_\rho$ and $S_\rho$, R.H.S. of equation (5) will not be convergent but Taylor expansion of $\ln\{\rho/(Z_\rho S_\rho)\}$ will converge yielding

$$\rho(x)dx = Z_\rho S_\rho \exp(-\Sigma_{i=0}^{\infty} a_i x^i)dx \qquad (6)$$

If $\rho(x)dx$ has finite discontinuities or real or imaginary or zeros or singularities in *[a,b]* at a finite number of interior points then *[a,b]* has to be subdivided into a finite number of intervals in each of which an expression of form (6) will hold. This form of density is non-negative in the entire domain unlike the Gram-Charlier series that seeks an improvement on the Gaussian. It also explains why central limit theorem yields a Gaussian. Equation (6) will ensure faster convergence of approximate solutions than a power series when differential equations are solved (Venkataraman, 1999). In the case of more than one variable there will be more constraint equations specified for example in two variable case by

$$\langle x^i y^j \rangle = c_{ij} \qquad (7)$$

which yields

$$\rho(x,y)dxdy = \exp[-\Sigma_{(i,j)=(0,0)}^{\infty} a_{ij} x^i y^j]dxdy \qquad (8)$$

where $(x,y) \in D \subset R$ and $\rho$ does not vanish anywhere in $D$. Coefficients $a_{ij}$ are derived from

$$\left\langle \frac{-\partial}{\partial a_{ij}}\{\ln\rho(x)\}\right\rangle = c_{ij} \tag{9}$$

Vide the discussion in appendix 1. Probability densities of the form (6) for one variable could be derived from information theory if the latter is redefined as

$$I = \langle \ln\{\rho/(Z_\rho S_\rho)\}\rangle \tag{10}$$

### III. Quantum Mechanical Equation Consistent with Information Theory :

Since information given by equation (10) yields a probability density that allows zeros and singularities at end points of *[a,b]* and that $\rho(x)dx$ being given by $|\psi|^2 dx$ in quantum mechanics where symbol $\psi$ is the state vector, zeros of $\Psi$ can also be in the interior of the domain, the following variational formulation is proposed:

Minimize *I* given by (10) subject to $\langle H\rangle$ being constant or vice versa which yields

$$-1/2 \, d^2\Psi_n/dx^2 + V(x)\Psi_n = \lambda_n[1 + \ln\{\Psi_n^*\Psi_n/(Z_n^*Z_n)\}]\Psi_n \tag{11}$$

in the place of Schroedinger equation, where $Z_n$ represents the zeros of $\psi_n$ in *D*, the domain of integration and $\psi_n \in L^2(D)$ and $\lambda_n$ is the Lagrange multiplier. These equations are nonlinear. The logarithmic term, depending upon the sign of the Lagrange multiplier, will increase/decrease the energy values. If normalisation is also introduced in equation (11) Schroedinger equation can be retrieved in the limit of the Lagrange multipltier corresponding to the logarithmic term tending to zero. The solutions of (11) are the normalised solutions. Since the above problem is the same as minimising energy subject to fixing the information, energy values will also be ordered as in the standard linear case.

In appendix 2 it is shown that the multipliers corresponding to orthogonality are zero and that equation (11) yields the unique, complete set of solutions in the appropriate Hilbert space.

The reader who recognizes the importance of using all moments as constraints in deriving equation (3) using (2) and (1) might ask why $\langle H^i\rangle$ for *i>1* have not been concluded in the variational formulation. At present the linear equation

$$[H + \beta H^2]\Psi = \lambda\Psi \tag{12}$$

arising from specifying $\langle H^2\rangle$ is being studied. Inclusion of the latter will yield correction terms. There are papers in literature where the logarithmic non-linear terms are introduced and the equations studied with different aims (Doebner et al, 1999, Weinberg, 1989).

### IV One Dimensional Harmonic Oscillator:

Equations (11) reduce to

$$-1/2 \, d^2\Psi_n/dx^2 + x^2/2\,\Psi_n = \lambda_n[1 + \ln\{\Psi_n^*\Psi_n/(Z_n^*Z_n)\}]\Psi_n \tag{13}$$

where the zeros of $\psi_n$ are represented by a polynomial $P_n$ of degree *n*. Starting from

$$\Psi_n(x) = x^k P_n(x) \exp[-\sum_{i=0}^{\infty} a_{2i} x^{2i}] \qquad (14)$$

and taking recourse to the differential equation it can be shown that

$$\Psi_n(x) = H_n(\sqrt{2\beta_n} x) \exp[-\alpha_n - \beta_n x^2] \qquad (15a)$$

$$k = 0 \text{ or } 1 \qquad (15b)$$

$$\lambda_n = (4\beta_n^2 - 1)/4\beta_n \qquad (15c)$$

$$\beta_n^2 = [2\alpha_n - 1]/[8(n + k + \alpha_n)] \qquad (15d)$$

$$\exp(2\alpha_n) = \sqrt{\Pi} 2^n n! / \sqrt{2\beta_n} \qquad (16)$$

where $H_n$'s are Hermite polynomials. The solutions to (16) being unique $\psi_n$**'s** given by (15) are unique. One can alternatively prove that the solutions are complete using the standard basis in $L^2 \circledR$: If $\psi \varepsilon L^2 \circledR$ then it can be written as a linear combination of the vectors of the standard basis, the set of Hermite functions It is easy to show then the orthogonality of this with those given in (15a) implies it is a null vector. The energy values in the various states $\psi_n$ are given by

$$E_n = \lambda_1 [1 - 2\alpha_n - (2n+1)/2] \qquad (17)$$

The numerical values of various quantities are given below for two cases:

| N | $\alpha_n$ | $\beta_n$ | $\lambda_n$ | $E_n$ |
|---|---|---|---|---|
| 0 | 0.561903 | 0.165957 | -1.34046 | 0.836186 |
| 1 | 0.8846183 | 0.182575 | -1.18673 | 2.69296 |
| 2 | 1.483947 | 0.265717 | -0.675132 | 3.01642 |
| 3 | 2.374767 | 0.271151 | -0.650844 | 4.71831 |
| 4 | 3.3791495 | 0.312319 | -0.488143 | 5.00752 |
| 5 | 4.5328009 | 0.309387 | -0.498664 | 6.76468 |
| 6 | 5.7558755 | 0.334322 | -0.413460 | 7.03368 |
| 7 | 7.07846158 | 0.330258 | -0.426725 | 8.81483 |

Since $\lambda_1$'s are negative the energy values are higher than in the linear case though the spacing between energies is not constant..

**V Conclusion :** A modified form of information is proposed in equation (10) so that the probability density obtained from it tallies with a continuous Riemann integrable one that could have real or imaginary zeros or singularities at end points of *[a,b].* For the case of many variables the method of information theory can be used to derive joint probability density as in equations (7) and (8). Power series expansion, equivalent to Taylor series, for two variables is given in appendix for standard functions. To ensure consistency of quantum mechanics with most probable predictions, the new variational formulation (35), that yields a unique, complete orthonormal set of solutions is suggested . As the proof in appendix 2 shows to obtain the orthonormal set of solutions in the Hilbert space to nonlinear Schroedinger equation one has to take recourse to the variational method to obtain the other conditional minima.. The non-linear Euler-Lagrange equations have been solved analytically for one dimensional harmonic oscillator. The results could be compared with experimental results only when the problem is solved in $L^2(-l,l)$ as in reality the systems are always finite (Venkataraman, 1999) preferably after incorporation of $\langle H^i \rangle$ for at least $i=2$ in (11).

**Acknowledgment**

The author desires to place on record acknowledgment of enlightening discussions with D.P.Dewangan, N.N.Rao, J.C.Parikh, S.M. Roy, T.Sengadir, V.Singh and R.K.Varma and support from the Council of Scientific & Industrial Research and I.I.T.. Bombay.

**Appendix 1**

A probability density of the form given in equation (8) can be used when solutions to differential equations are sought for. Before deriving it in yet another way power series in several real variables is considered below.

It is well known (Courant et al 1989, Vol.1) that starting from

$$P_n(x) = \sum_{i=0}^{n} a_i x^i \qquad (18)$$

where

$$a_i = 1/i! [d^i/dx^i P(x)]_{(x=0)} \qquad (19)$$

Taylor and his pupils defined power series starting from

$$f(x) = \sum_{i=0}^{N} x^i/i! [d^i/dx^i f(x)]_{(x=0)} + R_N \qquad (20)$$

In the above equation it has to be shown that the remainder $R_N$ tends to zero to justify taking the upper limit to infinity.

Analogously if one begins from

$$P_N(x,y) = \sum_{i+j=0}^{N} a_{ij} x^i y^j \qquad (21)$$

and defines

$$a_{ij} = \frac{1}{i!j!} [\partial^{(i+j)}/\partial x^i \partial y^j P]_{(x=0, y=0)} \qquad (22)$$

one could show, since

$$(1+ax)^k = 1 + k(ax)/1! + k(k-1)(ax)^2/2! + \ldots \qquad (23)$$

is convergent for $|ax| < 1$ for $a$ and $k$ real

$$(1+xy)^k = 1 + k(xy)/1! + k(k-1)(xy)^2/2! + \ldots \ldots \quad (24)$$

is convergent for $|xy|<1$ or equivalently $r<1$ in polar coordinates, $(x,y) \in R^2$. Similarly since the power series for *exp(a x)* is convergent for all $x \in R$. where *a* is real

$$\exp(xy) = 1 + (xy)/1! + (xy)^2/2! + \ldots \ldots + (xy)^r/r! + \ldots \quad (25)$$

and the above series is convergent for all $(x,y) \in R^2$. Similarly the power series expression for logarithm could also be derived to arrive at equation (8). Taylor expansion for two variables is defined by

$$f(x,y) = \sum_{(i,j)=(0,0)}^{N} a_{ij} x^i y^j + R_N \quad (26)$$

The coefficients in (24) and (25) and (26) are precisely those derived from an equation analogous to (22). Hence power series in two variables can be defined through

$$f(x,y) = \sum_{(i,j)=(0,0)}^{\infty} a_{ij} x^i y^j \quad (27)$$

where

$$a_{ij} = \frac{1}{i! \, j!} [\partial^{(i+j)}/\partial x^i \partial y^j f]_{(x=0, y=0)} \quad (28)$$

It is easier to consider convergence questions for functions in equations (8), (24), (25) and (28) of several variables using polar coordinates. Equation (27) is the limiting form of (26) and gives a more natural expression for $a_{ij}$ 's. Maxima and minima could also be found easily solving just one equation in the radial coordinate and checking the signature of the second derivative in that coordinate (R.P.Venkataraman 1999).

## APPENDIX 2

The equation

$$H\Psi_0 = \lambda_{n0} \Psi_0 \quad (29)$$

corresponds to finding the lowest eigenvalue of *H* subject to the normalisation

$$\int_D d\Gamma \Psi_0^* \Psi_0 = 1 \quad (30)$$

in which *D* is the domain of integration and $d\Gamma$ is the infinitesimal volume element. The equation for $\psi_1$ arising out of the additional constraint of orthogonalisation with respect to $\psi_0$ is

$$H\Psi_1 = \lambda_{11} \Psi_1 + \mu_0 \Psi_0 \quad (31)$$

where $\mu_0$ is the Lagrange multiplier corresponding to the condition of orthogonality. It can be shown that $\mu_0$ is zero using self adjointness of *H* and the orthogonality condition. Thus the eigenvalue $\lambda_1$ is the next conditional minimum. For any *n* it can be shown that

$$H\Psi_n = \lambda_{nn} \Psi_n \quad (32)$$

because the Lagrange multipliers corresponding to the orthogonality with $\psi_k$ for all $k<n$ are zero and that the eigenvalues are ordered. The purpose of the above is to show that the solution of variational problems are unique and that equation (32) is a special case in which the system of equations gets reduced to one.

Starting with a variational formulation consistent with information theory the system of nonlinear Euler-Lagrange equations

$$-1/2\, d^2\Psi_n/dx^2 + V(x)\Psi_n = \lambda_n[1+\ln\{\Psi_n^*\Psi_n/(Z_n^*Z_n)\}]\Psi_n \quad (33)$$

where $\lambda_n$ is the Lagrange multiplier and $Z_n$ represents the zeros of $\psi_n$ was derived by minimising

$$I = \langle\, \ln\{\Psi_n^*\Psi_n/(Z_n^*Z_n)\}\, \rangle \quad (34)$$

subject to $\langle H \rangle$ being constant or vice versa. For any variation $\delta\psi_n^*$ in $\psi_n^*$

$$\int_D d\Gamma(\Psi_n^*+\delta\Psi_n^*)[-\ln\{(\Psi_n^*+\delta\Psi_n^*)\Psi_n/(Z_n^*Z_n)\}+(1/\lambda_n)H]\Psi_n$$

$$- \int_D d\Gamma\,\Psi_n^*[-\ln\{\Psi_n^*\Psi_n/(Z_n^*Z_n)\}+(1/\lambda_n)H]\Psi_n = 0 \quad (35)$$

which yields equation (33) for $\psi_n$. It shall now be proved that the Lagrange multipliers corresponding to orthogonality are zero. Choosing $n=1$ one obtains introducing orthogonality with $\psi_0$

$$-1/2\, d^2\Psi_n/dx^2 + V(x)\Psi_n = \lambda_n[1+\ln\{\Psi_n^*\Psi_n/(Z_n^*Z_n)\}]\Psi_n + \mu_0\Psi_0 \quad (36)$$

Multiplying the above equation by $\psi_0^*$ and integrating, one obtains for the multiplier

$$\mu_0\lambda_1 \int_D d\Gamma[\delta\Psi_1\delta\Psi_0]\Psi_0^*\Psi_0 = \int_D d\Gamma[\delta\Psi_1\delta\Psi_0]\lambda_0\ln\{(\Psi_0^*\Psi_0)/(Z_0^*Z_0)\}\Psi_0\Psi_1$$

$$- \int_D d\Gamma[\delta\Psi_1\delta\Psi_0]\lambda_1\ln\{(\Psi_1^*\Psi_1)/(Z_1^*Z_1)\}\Psi_1\Psi_0^* \quad (37)$$

where all the terms have been multiplied by the factor in square brackets to facilitate integration by parts. Using $\psi_0\delta\psi_0=\delta(\psi_0^2/2)$  $\psi_1\delta\psi_1=\delta(\psi_1^2/2)$ one obtains on using the finiteness of the logarithmic term and the boundary conditions of the solutions the value of
$\mu_0$ to be zero. To prove the multipliers corresponding to orthogonality for all $\psi_n$ are zero one assumes the result true for $k=0$ to $(n-2)$ and proves as above that the multiplier corresponding to $k=(n-1)$ is zero. If the nonlinearity is of the form $|\psi 2|^t$ then it can be shown easily the multipliers corresponding to orthogonality are nonzero.

To prove that the solutions of (35) are unique, let the contrary be assumed. Let $\psi_1$ and $\psi_2$ be two normalised solutions corresponding to the given value of $\lambda$. Then the energy values of these two vectors could be found. It is easy to show as above that

$$E_1(\lambda) - E_2(\lambda) = 0 \quad (38)$$

which is a contradiction.

As mentioned in the text, this formulation gives the solutions corresponding to the complete set of conditional minima of the energy functional and hence $E_0 < E_1 ... < E_i ...$. To prove that this orthonormal set is complete let the contrary be assumed. Let $\psi \in L^2(D)$ be a nonzero vector and let $\{\varphi_i\}$ a basis in that space such that $\{\varphi^{(1)}\} \in C^2(D)$. Then $\psi$ can be written as a linear combination of the latter and the expectation value of the Hamiltonian operator in this state can be found. Then this value $E$ must lie between two values of the energies, say $E_i$ and $E_{(i+1)}$. This is a contradiction as $E_{(i+1)}$ is the next conditional minimum after $E_i$. Hence the solutions corresponding to the complete set of minima of the energy functional must be complete in the corresponding Hilbert space.